\documentclass[twocolumn,showpacs,preprintnumbers,amsmath,amssymb]{revtex4}

\allowdisplaybreaks
\allowdisplaybreaks[4]
\usepackage{CJK}
\usepackage{graphicx}
\usepackage{dcolumn}
\usepackage{bm}
\usepackage[dvipdfm,
            pdfstartview=FitH,
            CJKbookmarks=true,
            bookmarksnumbered=true,
            bookmarksopen=true,
            colorlinks,
            pdfborder=001,
            linkcolor=blue,
            anchorcolor=blue,
            citecolor=blue
            ]{hyperref}

\begin{document}

\begin{CJK}{GBK}{song}

\title{Chiral geometry in multiple chiral doublet bands}

\author{Hao Zhang}
 \affiliation{
State Key Laboratory of Nuclear Physics and Technology, School of Physics, Peking University,
 Beijing 100871, People's Republic of China}
\author{Qibo Chen}
 \email{qbchen@pku.edu.cn}
\affiliation{
State Key Laboratory of Nuclear Physics and Technology, School of Physics, Peking University,
 Beijing 100871, People's Republic of China}

\date{\today}

\begin{abstract}

The chiral geometry of the multiple chiral doublet bands with identical configuration is discussed for different triaxial deformation parameters $\gamma$ in the particle rotor model with $\pi h_{11/2}\otimes \nu h_{11/2}^{-1}$. The energy spectra, electromagnetic transition probabilities $B(M1)$ and $B(E2)$, angular momenta, and $K$-distributions are studied. It is demonstrated that the chirality still remains not only in the yrast and yrare bands, but also in the two higher excited bands when $\gamma$ deviates from $30^{\circ}$. The chiral geometry relies significantly on $\gamma$, and the chiral geometry of the two higher excited partner bands is not as good as that of the yrast and yrare doublet bands.

\end{abstract}

\pacs{21.60.Ev, 21.10.Re, 23.20.Lv}
\maketitle

\section{Introduction}

Spontaneous chiral symmetry breaking phenomenon captures significant attention nowadays. Since the chirality of a triaxial atomic nucleus was first predicted in 1997 by Frauendorf and Meng~\cite{Frauendorf1997NPA} and identified in 2001 by Starosta~{\it{et al.}}~\cite{Starosta2001PRL}, plenty of achievements have been made.

Experimentally, more than 30 candidate chiral nuclei have been found at $A\sim 80$~\cite{S.Y.Wang2011PLB, Koike2011IJMPE},
$100$~\cite{Vaman2004PRL, Joshi2004PLB, Timar2004PLB, Alcantara-Nunez2004PRC,
Y.X.Luo2004PRC, Joshi2005EPJA, S.J.Zhu2005EPJA, Timar2006PRC, Joshi2007PRL,
Timar2007PRC, Suzuki2008PRC, Y.X.Luo2009PLB, Sethi2013PLB, Tonev2014PRL,
Lieder2014PRL, Rather2014PRL},
$130$~\cite{Starosta2001PRL, Hartley2001PRC, Hecht2001PRC,
Koike2001PRC, Bark2001NPA, Starosta2002PRC,
Mergel2002EPJA, X.F.Li2002CPL, S.Zhu2003PRL, Hecht2003PRC, Koike2003PRC,
Rainovski2003PRC, Roberts2003PRC, Rainovski2003JPG, Simons2005JPG,
Grodner2006PRL, Petrache2006PRL, Tonev2006PRL, S.Y.Wang2006PRC,
Mukhopadhyay2007PRL, Tonev2007PRC, Y.X.Zhao2009CPL, Grodner2011PLB,
Timar2011PRC, K.Y.Ma2012PRC, Petrache2012PRC, Kuti2013PRC}
and $190$~\cite{Balabanski2004PRC, Lawrie2008PRC,
Lawrie2010EPJA, Masiteng2013PLB, Masiteng2014EPJA} mass regions.
More details see, e.g., Ref.~\cite{J.Meng2008MPLA, B.Qi2009thesis, J.Meng2010JPG,
J.Meng2014IJMPE, Bark2014IJMPE, Q.B.Chen2015thesis}.

Theoretically, many approaches have been come up to investigate nuclear chirality, such as the particle rotor model (PRM)~\cite{Frauendorf1997NPA, Starosta2002PRC,  Koike2003PRC, J.Peng2003PRC, J.Peng2003CPL,
Koike2004PRL, S.Q.Zhang2007PRC, S.Y.Wang2007PRC, Higashiyama2007EPJA, S.Y.Wang2007CPL, S.Y.Wang2008PRC,
S.Y.Wang2008CPC, B.Qi2009PLB, B.Qi2009PRC, S.Y.Wang2009CPL, S.Y.Wang2009CPC, Lawrie2010PLB,
Rohozinski2011EPJA, Rohozinski2011IJMPE, B.Qi2011PRC, B.Qi2011CPL, Shirinda2012EPJA},
the tilted axis cranking model (TAC)~\cite{Frauendorf1997NPA, Dimitrov2000PRL, Olbratowski2004PRL,
Olbratowski2006PRC}, the tilted axis cranking plus random phase
approximation (TAC+RPA)~\cite{Mukhopadhyay2007PRL, Almehed2011PRC}, the interacting boson fermion-fermion
model (IBFFM)~\cite{Brant2004PRC, Tonev2006PRL, Tonev2007PRC, Brant2008PRC, Brant2009PRC, Ganev2010PRC},
pair truncated shell model (PTSM)~\cite{Higashiyama2005PRC, Yoshinaga2005JPG, Yoshinaga2006EPJA},
projected shell model (PSM)~\cite{Bhat2014PLB}. Recently, based on the TAC, a microscopical collective
Hamiltonian was constructed and applied to the unified description of chiral vibration and
rotation~\cite{Q.B.Chen2013PRC}.

Due to the successful description of nuclear global properties and exotic phenomena~\cite{Ring1996PPNP, Vretenar2005RR, J.Meng2006PPNP}, the covariant density function theory (CDFT) is introduced to investigate the triaxial deformation of chiral nuclei with various configurations. Based on adiabatic and configuration-fixed constrain triaxial CDFT, the possible existence of multiple chiral doublets (M$\chi$D), i.e., more than one pair of chiral doublet bands in one single nucleus, was proposed for odd-odd nucleus $^{106}$Rh~\cite{J.Meng2006PRC}.
Then more M$\chi$D nuclei were predicted in the other rhodium isotopes~\cite{J.Peng2008PRC, J.M.Yao2009PRC, J.Li2011PRC}. Later on, the first experimental evidence for the predicted M$\chi$D was reported in $^{133}$Ce~\cite{Ayangeakaa2013PRL}, and also possibly in $^{107}$Ag~\cite{B.Qi2013PRC}.

In contrast to the M$\chi$D in which different partner bands are of distinct triaxial deformations and configurations, the M$\chi$D phenomenon are expected also with identical configuration, i.e., not only the yarst and yrare bands but also higher excited bands might be chiral partner bands~\cite{Droste2009EPJA, Q.B.Chen2010PRC, Hamamoto2013PRC}. The energy spectra of M$\chi$D are first presented with rigid and soft cores~\cite{Droste2009EPJA}. Then the chiral geometry of M$\chi$D was examined for configuration $\pi h_{11/2}\otimes \nu h_{11/2}^{-1}$ with constant and spin depended variable moments of inertia at $\gamma=30^{\circ}$~\cite{Q.B.Chen2010PRC}. It was confirmed that the M$\chi$D indeed can exist in identical configuration. The chiral geometry of M$\chi$D has been further examined for the configuration $\pi g_{9/2}\otimes \nu h_{11/2}^{-1}$ at $\gamma=30^{\circ}, 20^{\circ}$~\cite{Hamamoto2013PRC}. Very recently, the evidence of this type of M$\chi$D was first observed in $^{103}$Rh~\cite{Kuti2014PRL}. This observation shows that the chiral geometry in nuclear can be robust against the increase of the intrinsic excitation energy.

As mentioned previous, the M$\chi$D with identical configuration has been investigated for the ideal chiral system, i.e., one $h_{11/2}$ proton particle and one $h_{11/2}$ neutron hole coupled to a rigid rotor with triaxial deformation $\gamma=30^{\circ}$~\cite{Q.B.Chen2010PRC}, and its chiral geometry is examined by analyzing the evolution of angular momentum. Thus, it is intriguing and necessary to study the chiral geometry of M$\chi$D with identical configuration in more general cases where $\gamma$ deviates $30^{\circ}$.

In this paper, the physical observables and chiral geometry of the M$\chi$D with identical configuration are discussed for different triaxial parameters $\gamma$ with $\pi h_{11/2}\otimes \nu h_{11/2}^{-1}$ in PRM. The paper is organized as follows. In Sec.~\ref{Sec2}, the theoretical framework of PRM is briefly introduced. The numerical details are presented in Sec.~\ref{Sec3}. In Sec.~\ref{Sec4}, the obtained energy spectra, electromagnetic transition probabilities $B(M1)$ and $B(E2)$, angular momenta, as well as $K$-distributions are shown and discussed in details. Finally, the summary is given in Sec.~\ref{Sec5}.

\section{Theoretical framework}\label{Sec2}

The detailed theoretical framework for PRM can be found in Ref.~\cite{Frauendorf1997NPA, J.Peng2003PRC, S.Q.Zhang2007PRC, B.Qi2009PLB, B.Qi2011PRC}. In this section, for completeness, some key formulas are presented. The Hamiltonian of PRM is
\begin{equation}\label{eq1}
\hat{H}_{\mathrm{PRM}}=\hat{H}_{\mathrm{coll}}+\hat{H}_{\mathrm{intr}},
\end{equation}
in which the collective Hamiltonian is
\begin{equation}
\hat{H}_{\mathrm{coll}}=\sum_{k}\frac{\hat{R}_{k}^2}{2\mathcal{J}_k}=\sum_{k}\frac{(\hat{I}_{k}-\hat{J}_{k})^2}{2\mathcal{J}_{k}},
\end{equation}
where the indices $k=1,2,3$ refer to the principal axes in body-fixed frame of reference, and $\hat{I}_k$, $\hat{R}_k$, $\hat{J}_k$ represent the angular momenta of the total nucleus, core, and valence nucleons. The moments of inertia along three principal axes depend on triaxial deformation parameter $\gamma$, $\mathcal{J}_k=\mathcal{J}_0 \sin^2(\gamma-2\pi k/3)$~\cite{Ring1980book}. The Hamiltonian of intrinsic nucleon $\hat{H}_{\mathrm{intr}}$ takes the single-$j$ shell Hamiltonian~\cite{Frauendorf1997NPA, J.Peng2003PRC, S.Q.Zhang2007PRC, B.Qi2009PLB, B.Qi2011PRC}.

The eigenstates of PRM are obtained by diagonalizing the Hamiltonian~(\ref{eq1}) at strong coupling basis ~\cite{J.Peng2003PRC}
\begin{align}
|IM\alpha \rangle &=\sqrt{\frac{1}{2(1+\delta_{K0})}} \Bigg\{\sum_{K,k_{\mathrm{p}},k_{\mathrm{n}}} C^{IK\alpha}_{k_{\mathrm{p}}k_{\mathrm{n}}} \Big[|IMKk_{\mathrm{p}}k_{\mathrm{n}}\alpha  \rangle \notag\\
&+ (-1)^{I-j_{\mathrm{p}}-j_{\mathrm{n}}} |IM-K-k_{\mathrm{p}}-k_{\mathrm{n}}\alpha \rangle \Big] \Bigg\},
\end{align}
where $|IMK\rangle$ is the Wigner $D$ function, $|k_pk_n\rangle$ is the product of the proton and neutron states, $C^{IK\alpha}_{k_{\mathrm{p}}k_{\mathrm{n}}}$ is the expansion coefficient. The angular momentum projections onto the quantization axis (3-) in the intrinsic frame and the $z$ axis in the laboratory frame are denoted by $K$ and $M$, respectively. The other quantum numbers are denoted by $\alpha$.

With the obtained eigenstates, the reduced transition probability can be calculated according to
\begin{equation}
B(\sigma \lambda, I'\alpha'\rightarrow I\alpha)=\sum_{M'M}\Big{|}\langle \mathrm{f},IM\alpha|\hat{T}_{\lambda\nu}|\mathrm{i},I'M'\alpha'\rangle \Big{|}^2,
\end{equation}
where $\sigma=E$ or $M$ indicates electric or magnetic transition, respectively, and $\lambda$ is the
rank of electric or magnetic transition operator from initial state $|\mathrm{i}\rangle$ to final state $|\mathrm{f}\rangle$. Thus, for reduced electric quadrupole transition probability
\begin{align}
& B(E2, I'\alpha'\rightarrow I\alpha) \notag\\
&\quad =\frac{5Q_0^2}{16\pi} \Bigg{|} \sum_{K,K'}^{k_{\mathrm{p}},k_{\mathrm{n}}} C^{IK\alpha}_{k_{\mathrm{p}}k_{\mathrm{n}}}C^{I'K'\alpha'}_{k'_{\mathrm{p}}k'_{\mathrm{n}}} \Big{[} \cos\gamma\langle IK20|I'K'\rangle \notag\\
& \qquad -\frac{\sin\gamma}{\sqrt{2}}\big{(} \langle IK22|I'K'\rangle + \langle IK2-2|I'K'\rangle \big{)} \Big{]} \Bigg{|}^2;
\end{align}
for reduced magnetic dipole transition probability
\begin{align}
& B(M1, I'\alpha'\rightarrow I\alpha)\notag\\
&\quad =\frac{3}{16\pi} \Bigg{|} \sum_{\mu,k_{\mathrm{p}},k_{\mathrm{n}},k'_{\mathrm{p}},k'_{\mathrm{n}}} \frac{1}{(1+\delta_{K'0})}\frac{1}{(1+\delta_{K0})}\notag\\
&\qquad \times C^{IK\alpha}_{k_{\mathrm{p}}k_{\mathrm{n}}}C^{I'K'\alpha'}_{k'_{\mathrm{p}}k'_{\mathrm{n}}} \Big{[} \langle IK1\mu|I'K'\rangle \langle k'_{\mathrm{p}}k'_{\mathrm{n}}|[ (g_{\mathrm{p}}-g_{\mathrm{R}})j_{\mathrm{p}\mu}\notag\\
&\qquad +(g_{\mathrm{n}}-g_{\mathrm{R}})j_{\mathrm{n}\mu}  ]|k_{\mathrm{p}}k_{\mathrm{n}}\rangle \notag \\
&\qquad +(-1)^{I-j_{\mathrm{p}}-j_{\mathrm{n}}}\langle I-K1\mu|I'K'\rangle \langle k'_{\mathrm{p}}k'_{\mathrm{n}}|[ (g_{\mathrm{p}}-g_{\mathrm{R}})j_{\mathrm{p}\mu}\notag\\
&\qquad+(g_{\mathrm{n}}-g_{\mathrm{R}})j_{\mathrm{n}\mu} ]|-k_{\mathrm{p}}-k_{\mathrm{n}} \rangle \Big{]}+\mathrm{sign} \Bigg{|}^2.
\end{align}
Here, $Q_0$ is intrinsic quadrupole moment; $g_R$, $g_p$, and $g_n$ are gyromagnetic ratios of rotor, proton, and neutron.

\section{Numerical details} \label{Sec3}

In Refs.~\cite{Frauendorf1997NPA} and \cite{Q.B.Chen2010PRC}, the PRM with a particle-like $h_{11/2}$ proton and a hole-like $h_{11/2}$ neutron for $\gamma=30^\circ$ is respectively used to study the chiral doublet bands and the M$\chi$D with identical configuration. In Ref.~\cite{B.Qi2009PRC}, the same system is discussed with different triaxial deformation parameters $\gamma=30^\circ$, $27^\circ$, $24^\circ$, $21^\circ$, $18^\circ$, and $15^\circ$ to examine the $B(M1)$ staggering as a fingerprint for chiral doublet bands.  In this paper, the same model will be used to examine the robustness of M$\chi$D with identical configuration with respect to the triaxiality of nucleus. For all the PRM calculations, we employ a symmetric particle-hole configuration $\pi h_{11/2}\otimes \nu h_{11/2}^{-1}$, quadrupole deformation with $\beta=0.25$, and the moment of inertia $\mathcal{J}=30~\textrm{MeV}^{-1}\hbar^2$. For the electromagnetic transition calculations, the empirical intrinsic quadrupole moment $Q_0=(3/\sqrt{5\pi}R_0^2Z\beta)=3.5~e\textrm{b}$, gyromagnetic ratios $g_R=Z/A=0.44$, and $g_p=1.21$, $g_n=-0.21$ are used~\cite{B.Qi2009PRC, Q.B.Chen2010PRC} in accordance with the mass region $A\sim 130$. The numerical details used here are in agreement with Ref.~\cite{B.Qi2009PRC}, and only $\beta$ is sightly different ($\beta=0.22$ in Ref.~\cite{B.Qi2009PRC}). Thus, the discussions of bands 1 and 2 presented in Ref.~\cite{B.Qi2009PRC} still hold true here, and we focus on the higher excited bands 3 and 4.

\section{Results and discussion} \label{Sec4}

\subsection{Energy spectra}

\begin{figure}[!ht]
\begin{center}
\includegraphics[width=8cm]{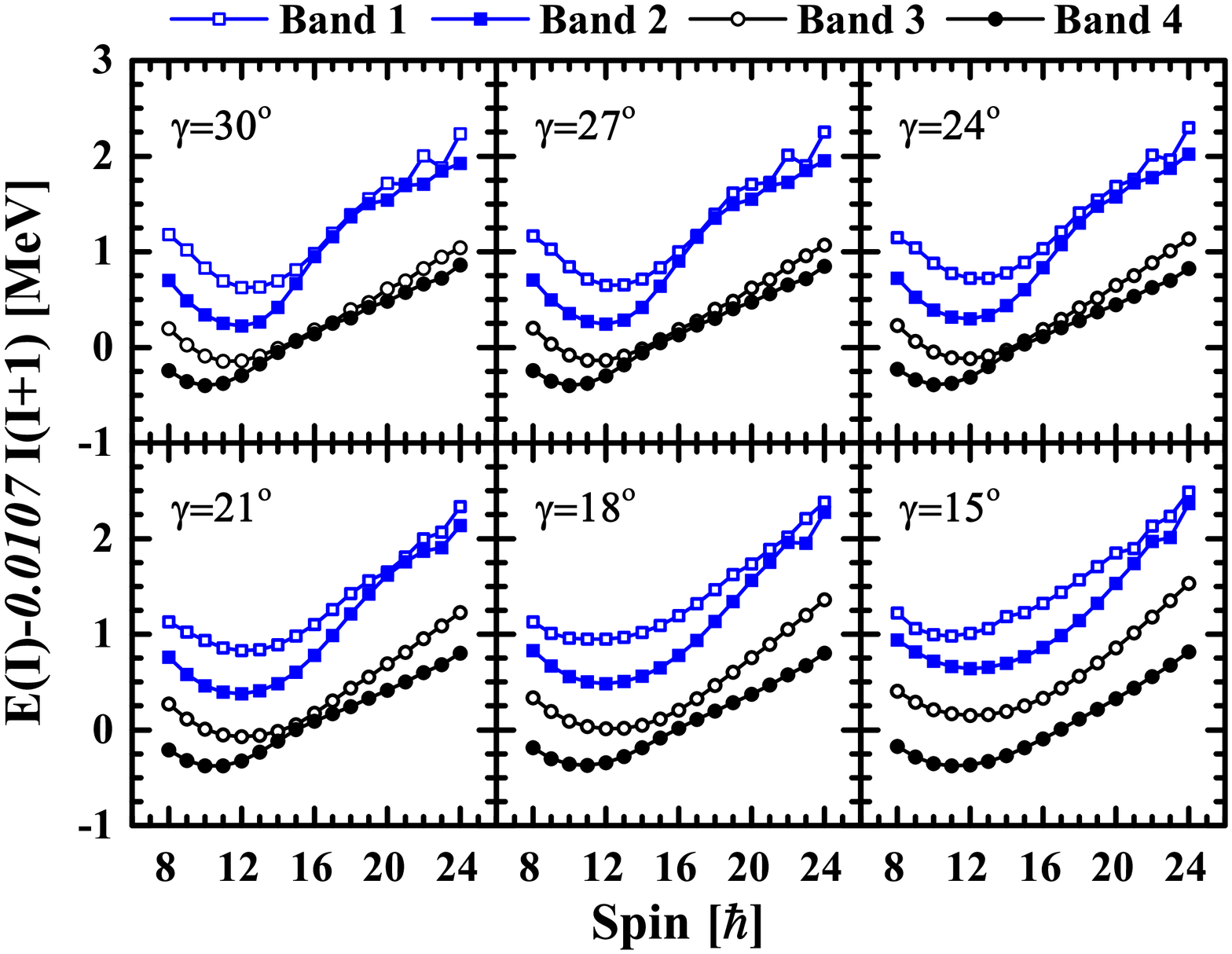}
\caption{(Color online) The energy spectra of the four lowest bands for the configuration $\pi h_{11/2}^1\otimes \nu h_{11/2}^{-1}$ with deformation parameter $\beta=0.25$ and moment of inertia $\mathcal{J}=30~\mathrm{MeV}^{-1}\hbar^2$ calculated by PRM with different triaxial deformation parameters $\gamma$. Energy of a rigid-rotor has been subtracted from the total energies. }\label{fig1}
\end{center}
\end{figure}

Fig.~\ref{fig1} shows the energy spectra of four lowest bands 1, 2, 3 and 4 calculated in PRM when triaxial deformation parameter $\gamma$ varies from $30^\circ$ to $15^\circ$. The discussions about the changing of the energy spectra with respect to $\gamma$ for bands 1 and 2 can be found in Ref.~\cite{B.Qi2009PRC}. For $\gamma=30^\circ$, it is difficult to distinguish the doublet bands 1 and 2. As $\gamma$ deviates from $30^\circ$, the energy difference between the bands 1 and 2 increases.

Similar behavior can be also found in doublet bands 3 and 4. For $\gamma=30^\circ$, analogous to bands 1 and 2, the energy of  bands 3 and 4 is identical when $16 \leq I \leq 19\hbar$. The energy differences between these two bands vanish, indicating that the static chirality appears. In addition, the energy differences between neighboring bands at low spin region ($8 \leq I \leq 12\hbar$) are approximately same, which may correspond to the chiral vibration~\cite{Mukhopadhyay2007PRL, B.Qi2009PLB}. When $\gamma$ gradually deviates from $30^\circ$, the degeneracy at $16  \leq I \leq 19\hbar$ is gradually removed. Especially when $\gamma=15^\circ$, the two energy difference between bands 3 and 4 is larger than $0.4~\textrm{MeV}$, suggesting the absence of static chirality. In contrast to the static chirality, the chiral vibration remains when $\gamma$ decreases.

\subsection{Electromagnetic transition probabilities}

\begin{figure}[!ht]
\begin{center}
\includegraphics[width=8cm]{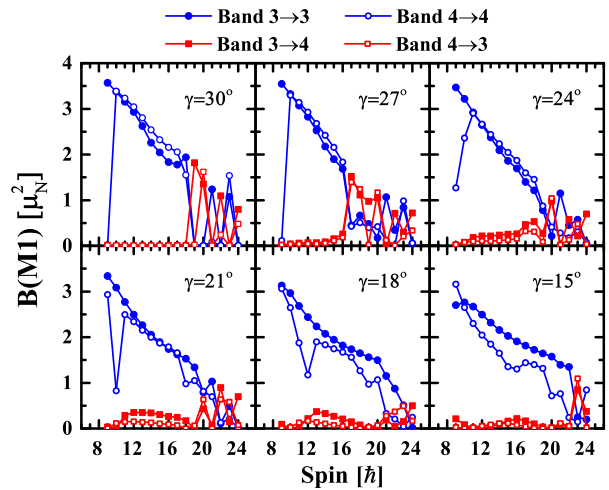}
\caption{(Color online) The intra- (band 3 $\rightarrow$ band 3 and band 4 $\rightarrow$ band 4) and inter-band (band 3 $\rightarrow$ band 4 and band 4 $\rightarrow$ band 3) reduced magnetic dipole transitions probabilities $B(M1)$ of bands 3 and 4 calculated by PRM with different triaxial deformation parameters $\gamma$.}\label{fig2}
\end{center}
\end{figure}

\begin{figure}[!ht]
\begin{center}
\includegraphics[width=8cm]{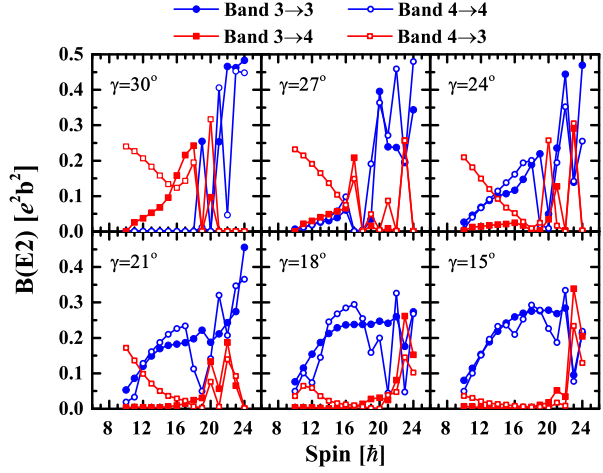}
\caption{(Color online) The intra- (band 3 $\rightarrow$ band 3 and band 4 $\rightarrow$ band 4) and inter-band (band 3 $\rightarrow$ band 4 and band 4 $\rightarrow$ band 3) reduced electric quadrupole transitions probabilities $B(E2)$ of bands 3 and 4 calculated by PRM with different triaxial deformation parameters $\gamma$.}\label{fig3}
\end{center}
\end{figure}

Besides the small energy difference, the similarity of the electromagnetic transition probabilities and the odd-even staggering of $B(M1)$ are remarkable characteristics for chiral doublet bands~\cite{Koike2004PRL, Petrache2006PRL, Tonev2006PRL, S.Y.Wang2007CPL, B.Qi2009PRC, Q.B.Chen2010PRC}. It has been demonstrated that for the yrast and yrare bands, the $B(M1)$ staggering is weak in the chiral vibration region, while it is strong in the static chirality region~\cite{B.Qi2009PRC}. Here, we discuss the electromagnetic transition probabilities for bands 3 and 4. The calculated  intra- and inter-band reduced magnetic dipole transitions probabilities $B(M1)$ and electric quadrupole transitions probabilities $B(E2)$ of bands 3 and 4 by PRM are respectively shown in Fig.~\ref{fig2} and~\ref{fig3} with different triaxial deformation parameters $\gamma$. Notice that the intra-band transition probabilities of the doublet bands exhibit similar tendency. The same phenomena also take place in inter-band transition probabilities. This suggests that the chiral doublet bands might remain in bands 3 and 4 when $\gamma$ deviates $30^\circ$.

As shown in Fig.~\ref{fig2}, for $\gamma=30^{\circ}$, when $I\leq 15\hbar$, the intra-band $B(M1)$ decreases gradually with spin, while the inter-band $M1$ transitions are forbidden. For $I>15\hbar$, a strong staggering between bands 3 and 4 can be clearly seen, which means there is an alternative suppression between the intra- and inter- transition as spin increases. This has been demonstrated in Ref.~\cite{Q.B.Chen2010PRC}. However, as $\gamma$ deviates from $30^{\circ}$, two obvious changes happen. One is that the staggering becomes weaker and weaker at high spin region when $\gamma$ varies from $27^{\circ}$ to $15^{\circ}$; the other is that the suppression of inter-band transitions is gradually relieved at low spin region. Compared with the first partner bands 1 and 2, for which the staggering when $\gamma=15^{\circ}$ completely disappears~\cite{B.Qi2009PRC}, the staggering of the second partner bands 3 and 4 still exists at the same spin internal.

Fig.~\ref{fig3} shows the electric quadrupole transitions probabilities $B(E2)$ between the second partner bands. It can be seen that for $\gamma=30^{\circ}$, when $I\leq 15\hbar$, the intra-band $E2$ transitions are forbidden, and the inter-band ones are allowed. When $I>15\hbar$, on the contrary, the intra-band $E2$ transition are allowed, and the inter-band $E2$ transitions are forbidden. As $\gamma$ deviates from $30^{\circ}$, the probabilities of both intra- and inter-transitions for $I>20\hbar$ are similar, which indicates that $B(E2)$ is irrelevant to $\gamma$ at high spin. At low spin ($I\leq 20\hbar$), the situation is inverse: the suppressed inter-bands transitions are gradually allowed, yet intra-bands are gradually forbidden. The similar conclusions for bands 1 and 2 has been derived in Ref.~\cite{B.Qi2009PRC}.

Therefore, from the discussions above, it can be found that for $I>18\hbar$, the odd-even staggering of $B(M1)$ will lead to the staggering of $B(M1)/B(E2)$ and $B(M1)_{\mathrm{in}}/B(M1)_{\mathrm{out}}$. When $\gamma$ gradually decreases, the amplitude of $B(M1)/B(E2)$ and $B(M1)_{\mathrm{in}}/B(M1)_{\mathrm{out}}$ staggering will decline because of the weakening of $B(M1)$ staggering.

\subsection{Angular momenta}

\begin{figure}[!ht]
\begin{center}
\includegraphics[width=7 cm]{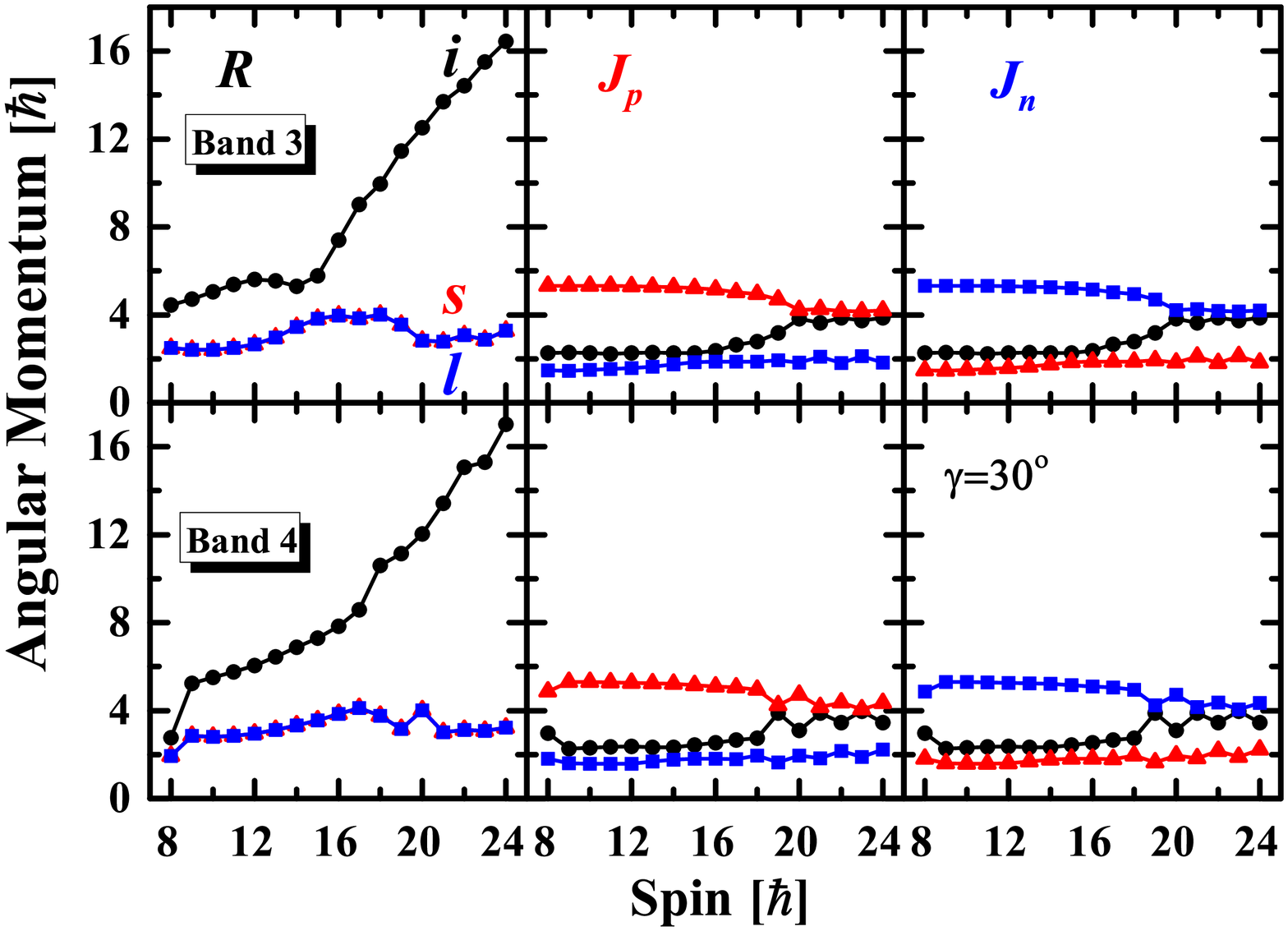}\\
\includegraphics[width=7 cm]{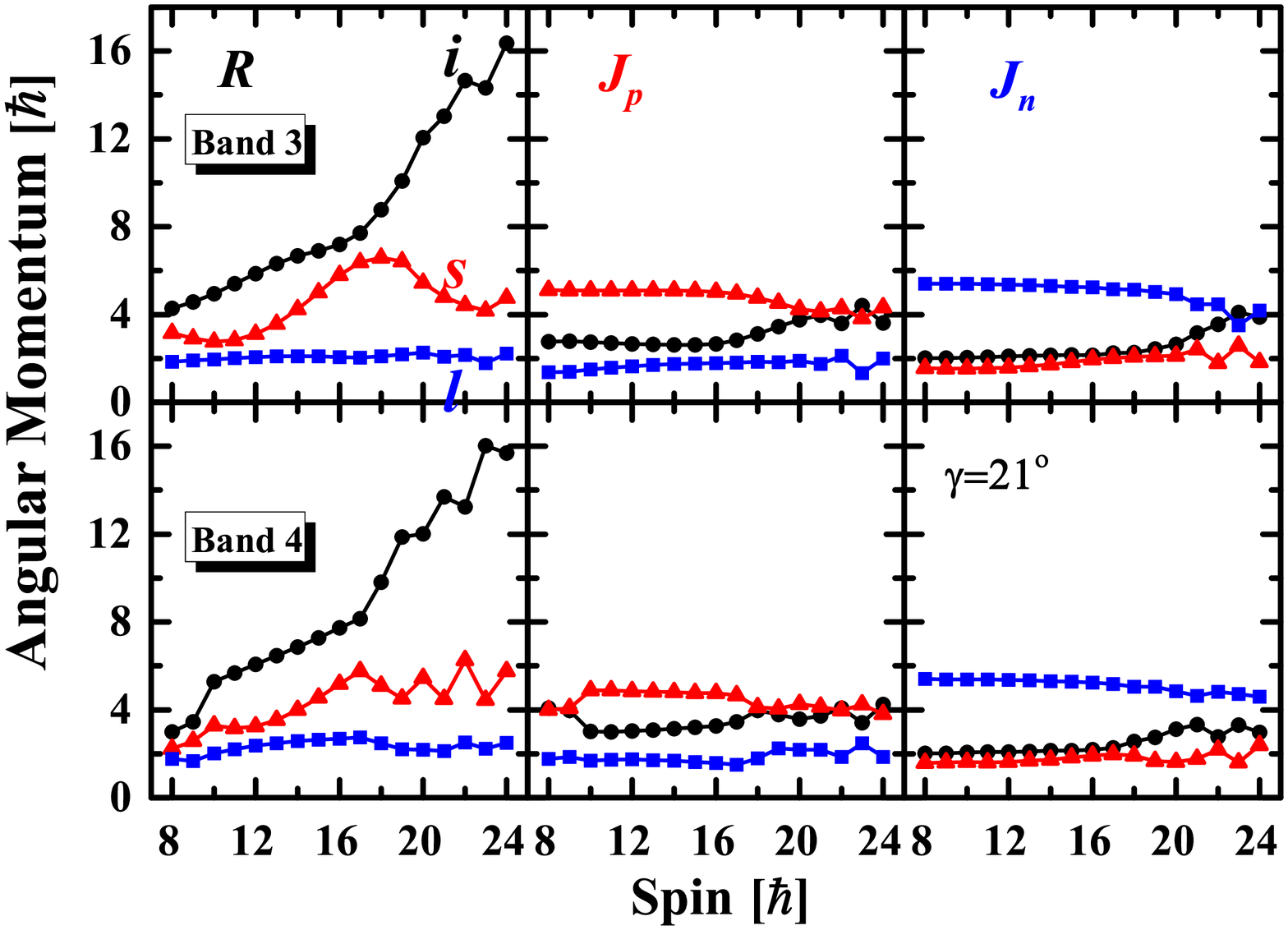}\\
\includegraphics[width=7 cm]{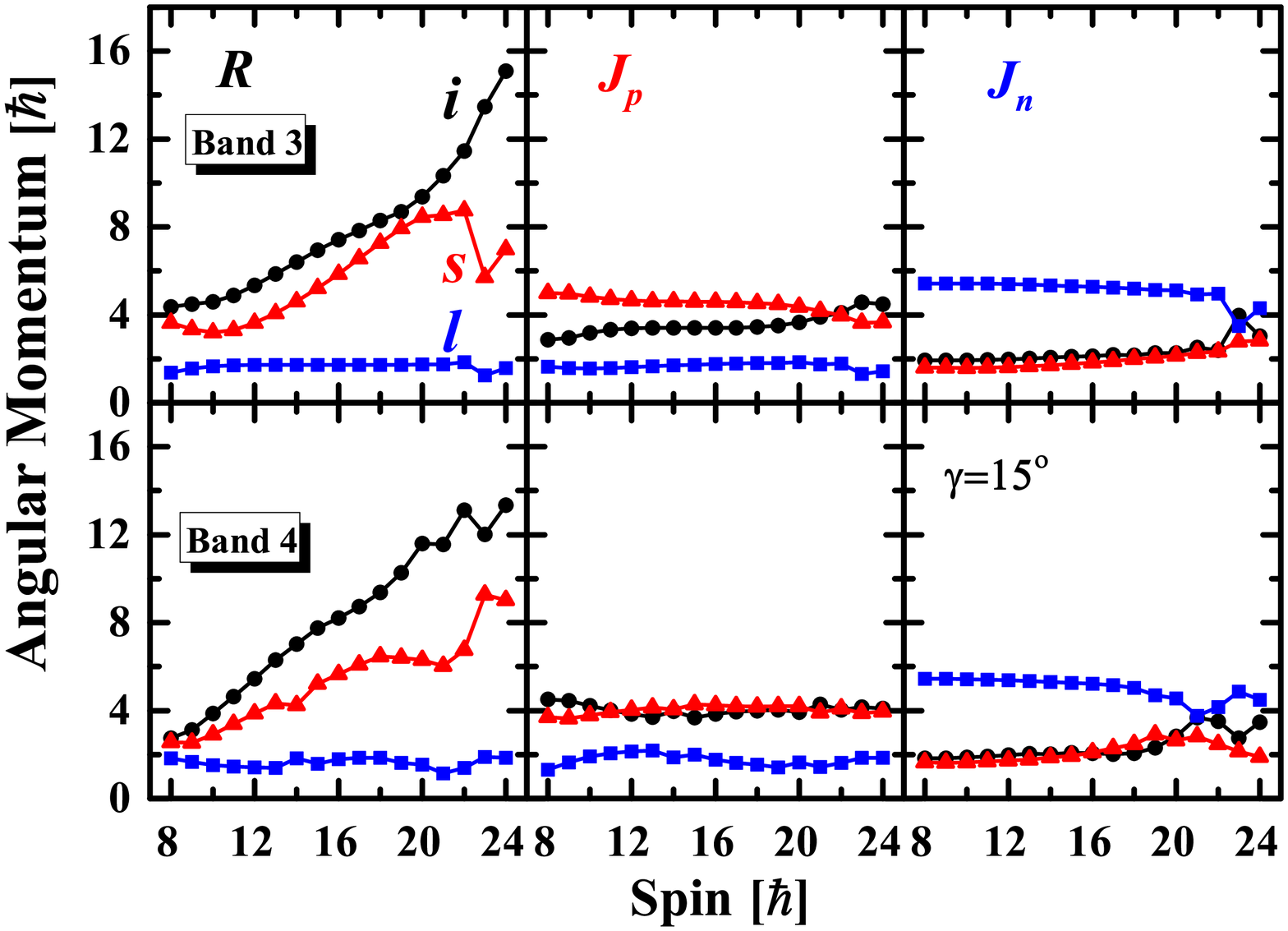}
\caption{(Color online) The expectation values of angular momenta of the core ($R_k=\sqrt{\langle \hat{R}_k^2 \rangle}$), the valence proton particle ($J_{\mathrm{p}k}=\sqrt{\langle \hat{J}_{\mathrm{p}k}^2 \rangle}$), and the valence neutron hole ($J_{\mathrm{n}k}=\sqrt{\langle \hat{J}_{\mathrm{n}k}^2 \rangle}$) along the intermediate ($\mathit{i}$), short($\mathit{s}$), and long ($\mathit{l}$) axes with the triaxial deformation parameter $\gamma=30^{\circ}, 21^{\circ}, 15^{\circ}$.}\label{fig4}
\end{center}
\end{figure}

An insight into chiral geometry will be revealed when it comes to the rms values of the angular momenta components of the core $R_k=\langle \hat{R}_k^2 \rangle^{1/2}$, the valence proton $J_{pk}=\langle \hat{j}_{pk}^2 \rangle^{1/2}$, and the valence neutron $J_{nk}=\langle \hat{j}_{nk}^2 \rangle^{1/2}$ ($k=1, 2, 3$). The results for bands 1 and 2 have already shown in Ref.~\cite{B.Qi2009PRC}, where the chiral geometry for the doublet bands is discussed. The obtained results for bands 3 and 4 at triaxial deformation parameter $\gamma=30^{\circ}, 21^{\circ}, 15^{\circ}$ are shown in Fig.~\ref{fig4}, in which $l, i, s$ correspond to long, intermediate and short axes. It is shown that for both partner bands, the core angular momentum mainly aligns along the $i$-axis due to its largest moment of inertia. The contribution of the valence proton particle mainly along the $s$-axis, and that of the valence neutron hole mainly aligns along the $l$-axis, for such orientations are favored by their interaction with the triaxial core~\cite{Frauendorf1997NPA}. With the total angular momentum increasing, $\bm{R}$ increases gradually, $\bm{J}_n$ and $\bm{J}_p$ move gradually toward the $i$-axis due to the Coriolis interaction, and the three angular momenta together form the chiral geometry of aplanar rotation.

From the left panel, it can be easily seen that when $\gamma=30^{\circ}$, the angular momenta of bands 3 and 4 are almost same, except at band head and $I=20\hbar$. They have the similar orientation. However, as $\gamma$ gradually deviates from $30^{\circ}$, this similarity disappears gradually. When $\gamma=21^{\circ}$, the differences between the partner bands become larger than those for $\gamma=30^{\circ}$, yet at $I=20,21,23\hbar$, two bands are still similar. As $\gamma$ continually decreases, such as $\gamma=15^{\circ}$, the orientation of angular momenta at $20 <I<24\hbar$ of the core, valence proton and neutron are different, corresponding to the large energy difference between bands 3 and 4 shown in Fig.~\ref{fig1}.

\subsection{$K$-distributions}

To better understand the mechanism of chiral geometry, the $K$-distributions, which reveal the probability distributions of the projection on three principle axes of the total angular momenta, are studied.

\begin{figure}[!ht]
\begin{center}
\includegraphics[width=5 cm]{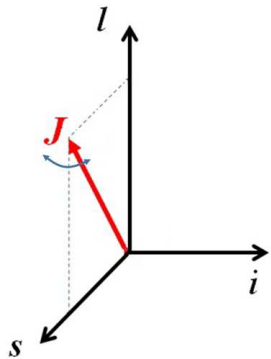}
\caption{(Color online) Diagrammatic sketch of chiral vibration of total angular momentum.}\label{fig7}
\end{center}
\end{figure}

\begin{figure}[!ht]
\begin{center}
\includegraphics[width=5 cm]{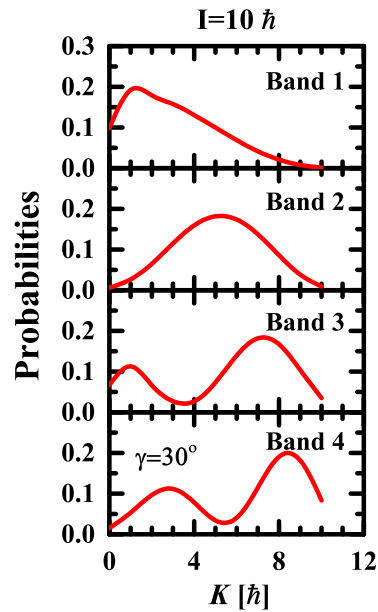}
\caption{(Color online) The $K$-distribution of four lowest bands 1, 2, 3, and 4
for $\gamma=30^{\circ}$ when $I=10\hbar$.}\label{fig8}
\end{center}
\end{figure}

As discussed above, at low spin, the valence proton particle and neutron hole contribute most to the total angular momentum. Thus the angular momentum mainly lies in $s$-$l$ plane, as diagrammatically shown in Fig.~\ref{fig7}. Due to the effect of quantal fluctuation, the total angular momentum will vibrate through $s$-$l$ plane. In this case, the chiral vibration will appear since the total angular momentum oscillates between the left-handed and right-handed systems~\cite{Mukhopadhyay2007PRL, B.Qi2009PLB}. To illustrate the vibrational character of this dynamical chirality procedure, Fig.~\ref{fig8} shows the $K$-distribution along the $i$-axis of four lowest energy bands for $\gamma=30^{\circ}$ at $I=10\hbar$. The $K$-distribution of bands 1 and 2 have been shown in Ref.~\cite{B.Qi2009PRC}, where it has been pointed out that bands 1 and 2 correspond to the zero- and one- phonon states with symmetric and antisymmetric wave functions on the $i$-axis. In the same way, bands 3 and 4 represent two- and three-phonon states with symmetric and antisymmetric wave functions. The chiral vibration character exhibits at low spin region can explain the similar spacing between the neighboring bands at the beginning of the M$\chi$D as shown in Fig.~\ref{fig1}.

\begin{figure}[!ht]
\begin{center}
\includegraphics[width=7 cm]{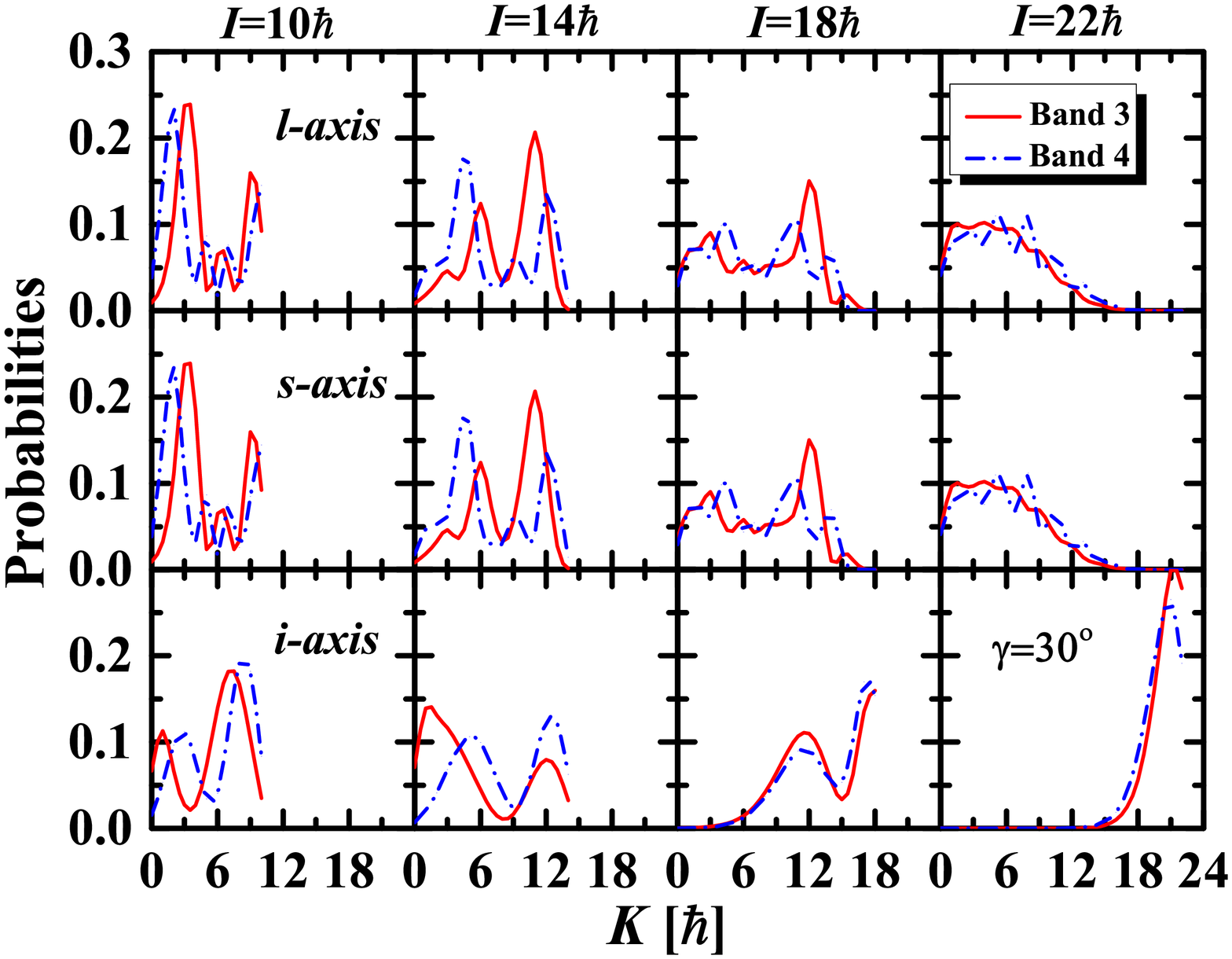}\\
\includegraphics[width=7 cm]{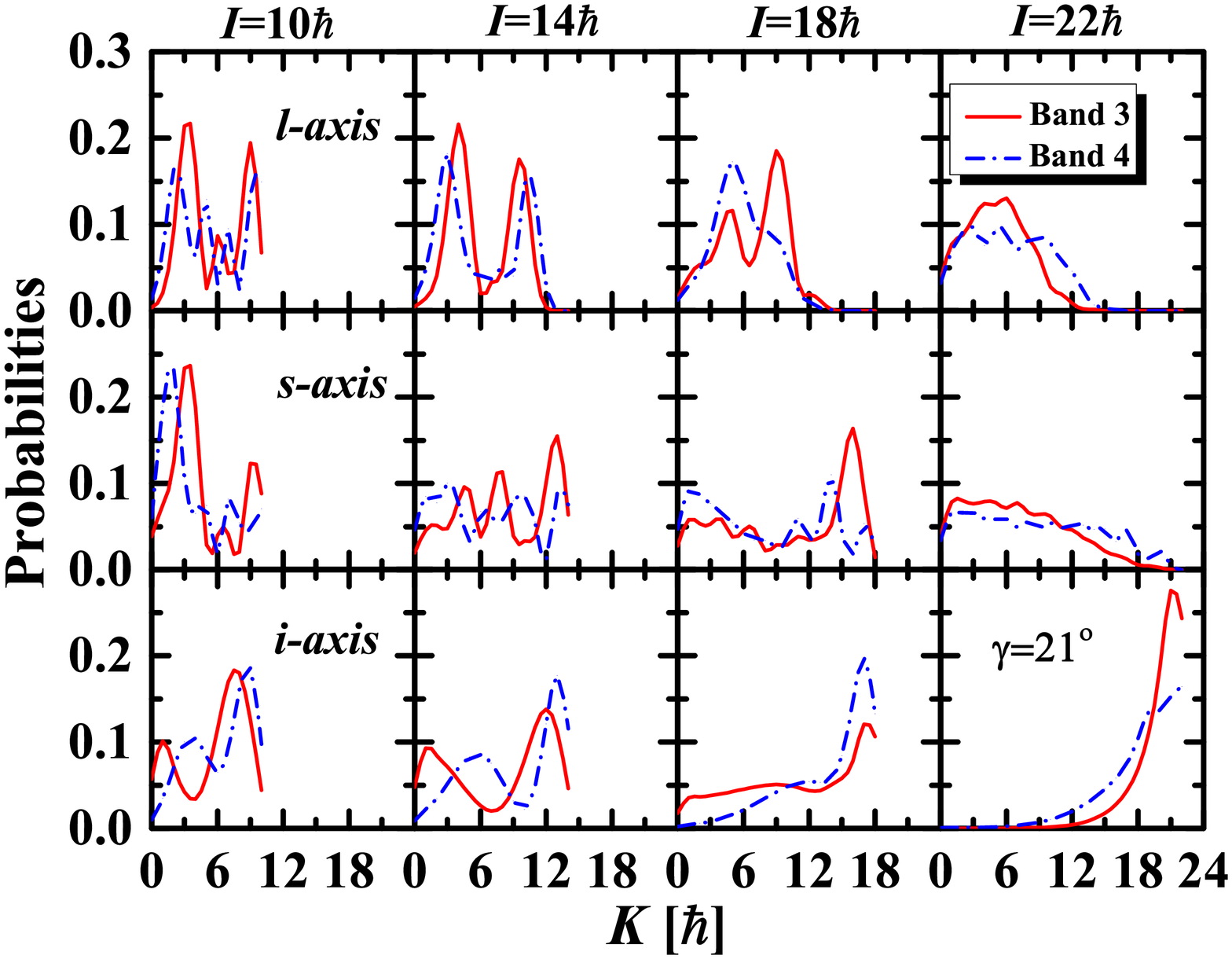}\\
\includegraphics[width=7 cm]{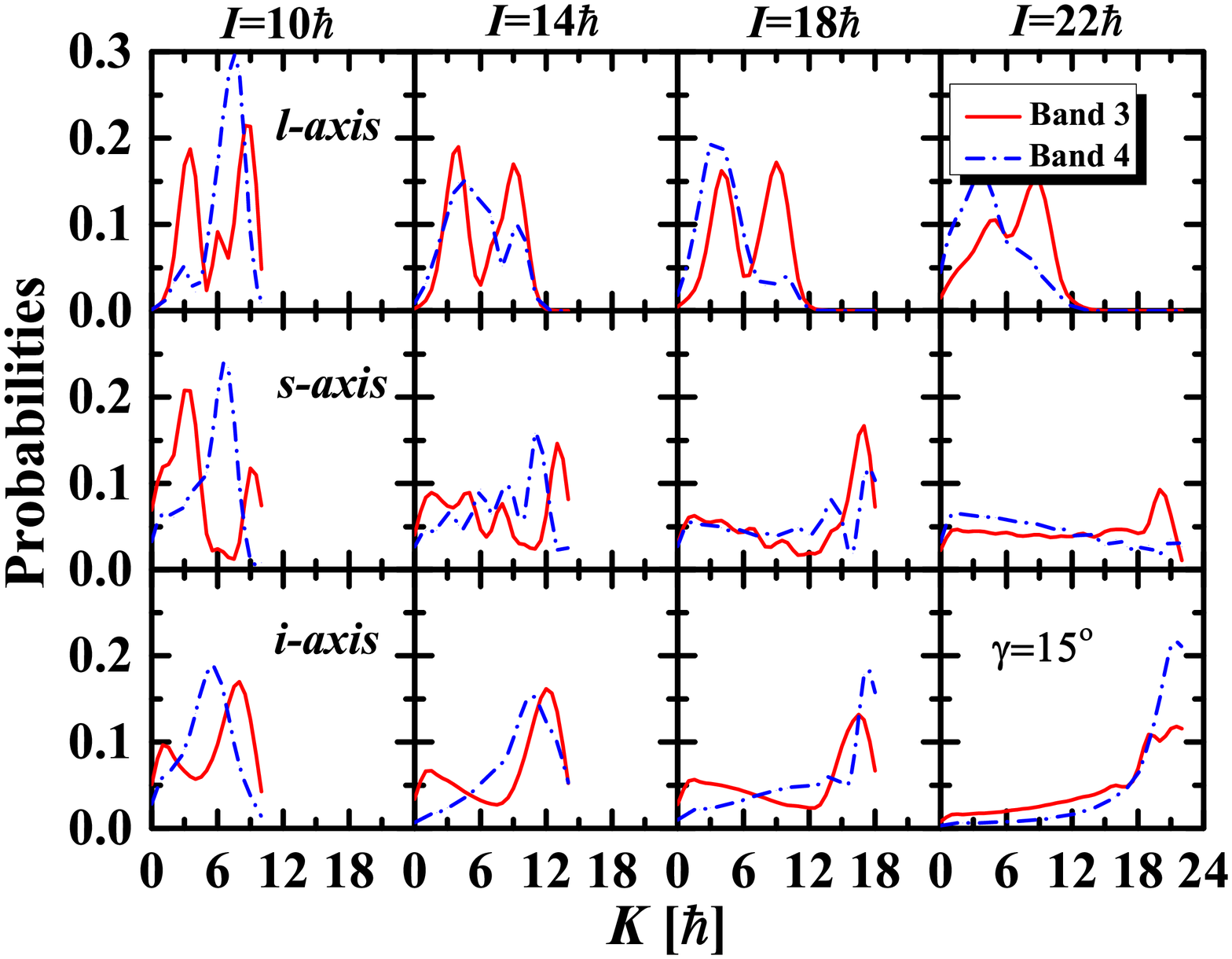}
 \caption{(Color online) The probabilities distributions of the projection of total angular momenta of bands 3 and 4 along the intermediate, short, and long axes when the parameter $\gamma=30^{\circ}, 21^{\circ}, 15^{\circ}$.}\label{fig5}
\end{center}
\end{figure}

The obtained results for bands 3 and 4 at triaxial deformation parameter $\gamma=30^{\circ}, 21^{\circ}, 15^{\circ}$ are shown in Fig.~\ref{fig5}. In the following discussions, we use $K_s$, $K_i$, and $K_l$ to represent the distributions on the short, intermediate, and long axis.

For $\gamma=30^{\circ}$, at higher spin region, such as $I=18, 22\hbar$, the $K$-distributions of the partner bands turn to be similar, implying the alteration from chiral vibration to static chirality~\cite{Q.B.Chen2010PRC}.

As $\gamma$ decreases, e.g., $\gamma=21^\circ$, the $K$-distribution of bands 3 and 4 when $I<20\hbar$ is similar to that when $\gamma=30^{\circ}$. But another mode of new type of chiral vibration is found when $I>20\hbar$. The maximum $K_s$ distribution at that internal appears near $K_s=0$ for band 3 and band 4, which corresponds to a chiral vibration through the $l$-$i$ plane.

When $\gamma=15^{\circ}$, two modes of chiral vibration remain at low and high spin, but the pure static chirality is hardly distinguishable. Thus, for spin $16 <I<20\hbar$, chiral vibration is mixed with static chirality, which may characterize energy spectra by the removal of degeneracy at $16 <I<20\hbar$.

\subsection{Energy spectra for higher excited states}

\begin{figure*}
\begin{center}
\includegraphics[width=6 cm]{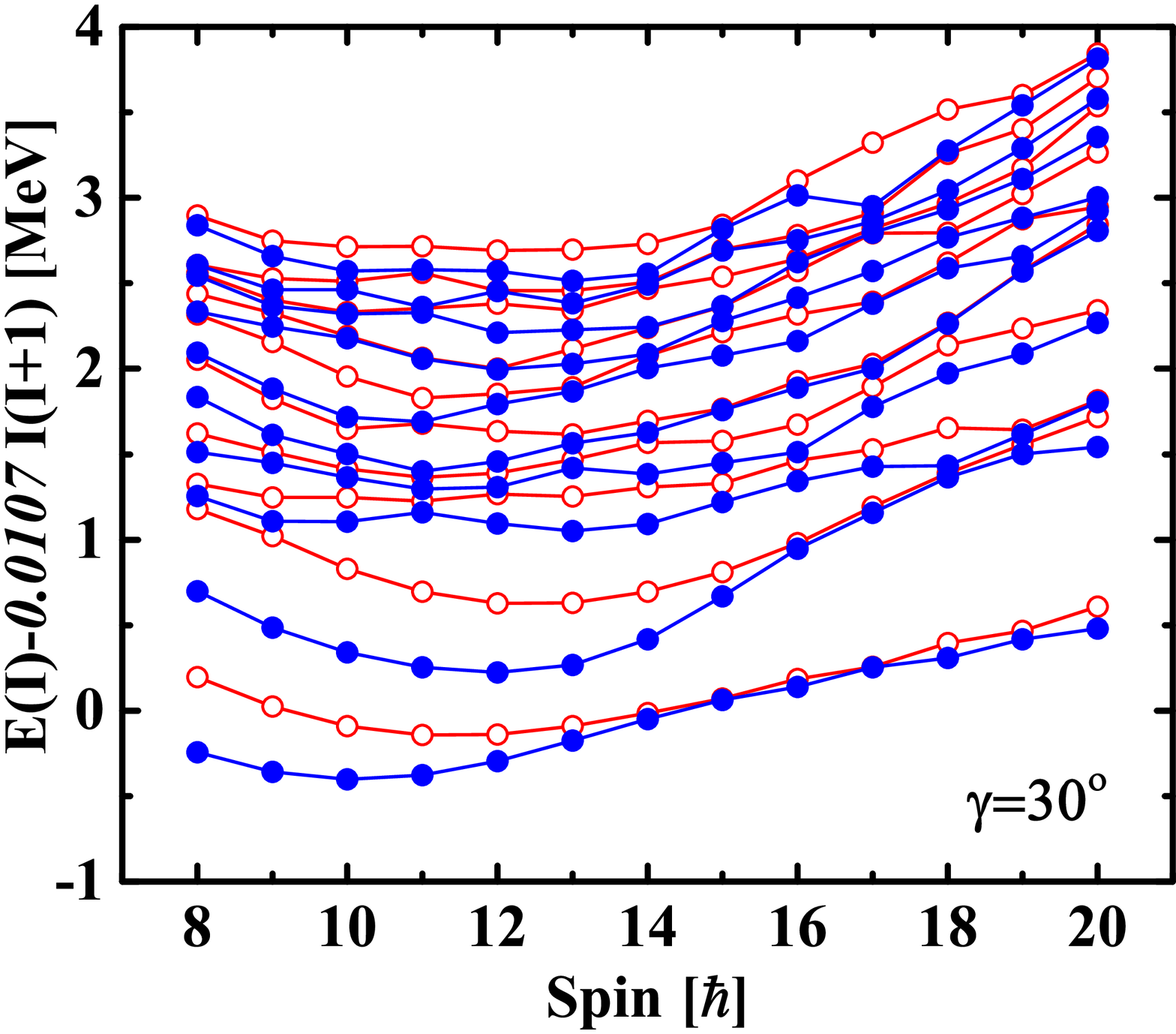}\quad\quad
\includegraphics[width=6.3 cm]{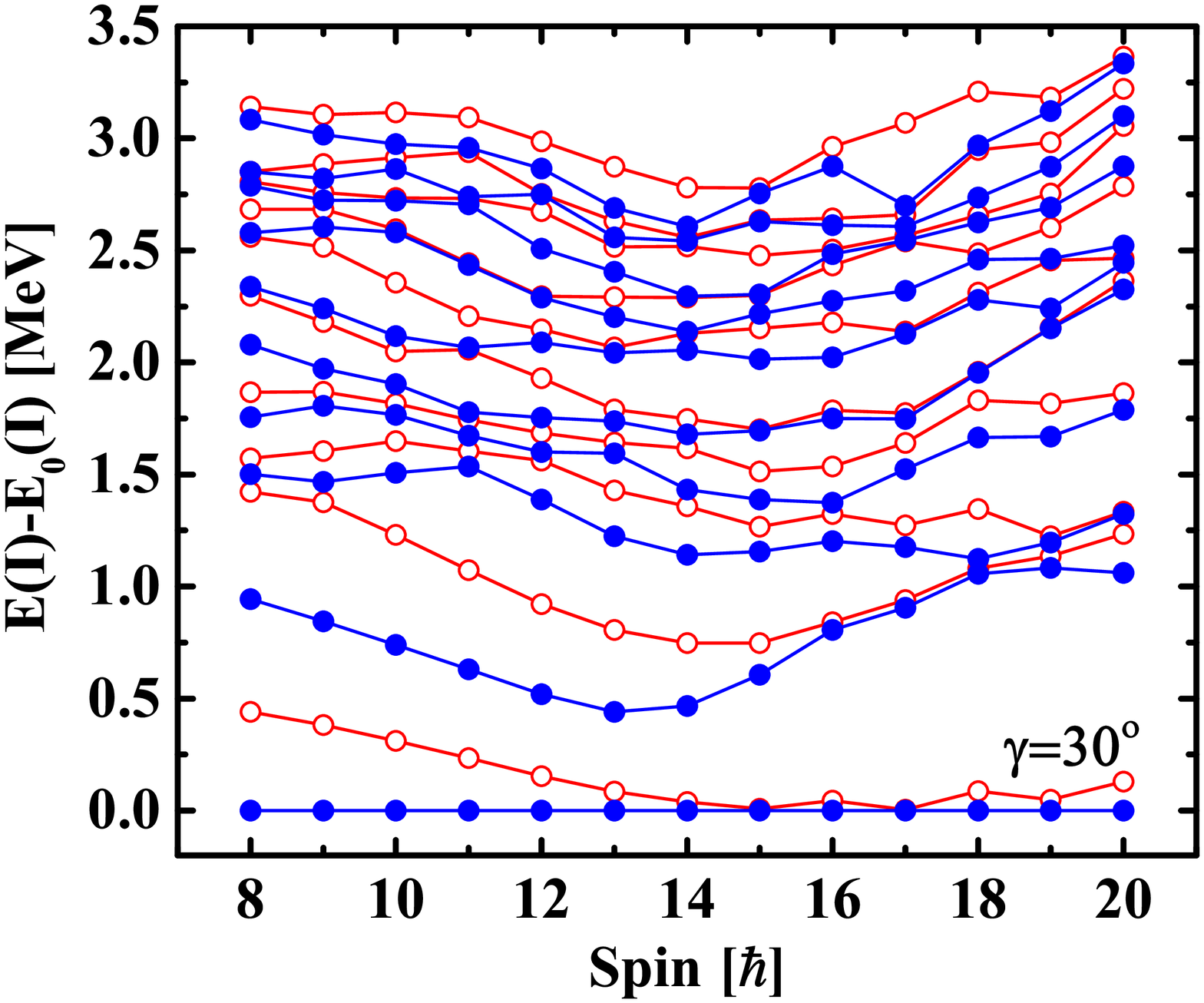}
\caption{(Color online) The energy states and their excitation energies with respect to yrast state obtained by PRM with triaxial deformation parameters $\gamma=30^\circ$. For each spin, the lowest twenty states are shown.}\label{fig9}
\end{center}
\end{figure*}

The discussion above gives the results of the physical observables and chiral geometry of the M$\chi$D with identical configuration for different triaxial parameters $\gamma$ with $\pi h_{11/2}\otimes \nu h_{11/2}^{-1}$. It has been shown that the chirality still remains not only in the yrast and yrare bands, but also in the two higher excited bands when $\gamma$ deviates from $30^{\circ}$. Further consideration of higher excited bands seems interesting. In Ref.~\cite{Q.B.Chen2010PRC, Q.B.Chen2013PRC}, the energy spectra of higher excited bands 5 and 6 have been shown at $\gamma=30^\circ$. It is found that bands 5 and 6 might also be a pair of chiral doublet bands. In Fig.~\ref{fig9}, the energy states as well as their excitation energies with respect to the yrast states obtained by PRM with triaxial deformation parameters $\gamma=30^\circ$ are shown as functions of spin. For each spin, the lowest twenty states are shown. It can be seen that in high excited bands, whose excitation energies are larger than 2.0 MeV, the situation turns to be complex. The level density is rather high and neighboring bands are too close to be distinguished. Thus it would be very difficult, if not impossible, to select the chiral doublet bands~\cite{Q.B.Chen2010PRC}. In fact, these states would fall into the regime of quasi-continuum spectra and the strong interactions between them evoke the emergence of rotation damping~\cite{Dossing1996PR, Bracco2002RPP}. This complexity can be explained by the PRM Hamiltonian~(\ref{eq1}), where the single particles, the rotor, and their coupling Hamiltonian are involved. If the single particles were not considered, it would turn to be a triaxial wobbling rotor Hamiltonian~\cite{Bohr1975}. At the spin region ($8\sim 20\hbar$) discussed above, the wobbling rotor has low angular momentum ($\sim 10\hbar$), which corresponds to a complicated wobbling energy spectrum~\cite{Bohr1975, Frauendorf2014PRC, Q.B.Chen2014PRC, W.X.Shi2015CPC}. With the coupling of the single particles, the occupation of proton and neutron at different orbitals further complicates the wobbling energy spectra of triaxial rotor.

In addition, the energy spectra of higher excited states with triaxiality parameters smaller than $\gamma=30^\circ$ are also examined. We note that the triaxial deformation has larger influence on the low lying states than high ones. When the triaxial deformation parameter varies from $30^\circ$ to $15^\circ$, the complexity of the high energy states still remain. Anyway, more efforts need to be paid to investigate the higher excited states.


\section{Summary}\label{Sec5}

In summary, the chiral geometry of the M$\chi$D with identical configuration is discussed for different triaxial deformation parameters $\gamma$ with $\pi h_{11/2}\otimes \nu h_{11/2}^{-1}$ in particle rotor model. The energy spectra, electromagnetic transition probabilities $B(M1)$ and $B(E2)$, angular momenta, and $K$-distributions are studied. The calculated results strongly suggest the chirality still remains in not only the yarst and yrare bands but also two higher excited bands when $\gamma$ deviates from $30^{\circ}$. The chiral geometry relies on $\gamma$, for the chiral vibration and static chirality exist in different spin internal when $\gamma$ changes. In addition, the chiral geometry of the higher energy partner bands is not as obvious as the ones with the lowest energy.
~~\\


\section*{Acknowledgements}

The authors are grateful to Professor Jie Meng, Professor Shuangquan Zhang,
and Wenxian Shi for fruitful discussions and critical reading of the manuscript.
The work of Hao Zhang was supported in part by the plan project of Beijing college
students's scientific research and entrepreneurial action.
This work was partly supported by the Major State 973 Program of China (Grant No. 2013CB834400),
the National Natural Science Foundation of China (Grants No. 11175002, No. 11335002,
No. 11375015, No. 11461141002), the National Fund for
Fostering Talents of Basic Science (NFFTBS) (Grant No. J1103206), the Research
Fund for the Doctoral Program of Higher Education (Grant No. 20110001110087).

\end{CJK}

\end{document}